# The high temperature mechanical properties and the correlated microstructure/ texture evolutions of a TWIP high entropy alloy

A. Nabizada[a], A. Zarei-Hanzaki[a, 1], H.R. Abedi[b], M.H. Barati[a], P. Asghari-Rad[c, d], H.S. Kim[c, d]

[a] Hot Deformation and Thermomechanical Processing Laboratory of High-Performance Engineering Materials, School of Metallurgy and Materials Engineering, College of Engineering, University of Tehran (UT), Tehran, Iran

[b] School of Metallurgy & Materials Engineering, Iran University of Science and Technology (IUST), Tehran, Iran

[c] Center for High Entropy Alloys, Pohang University of Science and Technology (POSTECH), Pohang, 37673, South Korea

[d] Department of Materials Science and Engineering, Pohang University of Science and Technology (POSTECH), Pohang, 37673, South Korea

**Abstract**

The present work deals with the microstructure and texture evolutions of a TWIP high entropy alloy at various deformation temperatures. Toward this end, the tensile tests were conducted at temperatures ranging from 298 to 873 K under the strain rate of $0.001 s^{-1}$. The experimented material exhibited an extraordinary room temperature work hardening behavior, in respect of both the instantaneous strain hardening exponent and the work hardening rate values. This resulted in an acceptable ultimate tensile strength /ductility balance and was justified considering the high activity of twinning and planar slip. Such work hardening capacity deteriorated at higher temperatures due to increasing the dislocation annihilation rate and the stacking fault energy values that suppressed twin formation and dislocation accumulation. However, the decrease in amplitude and extent of the work hardening region was insignificant compared with other single-phase high entropy alloys. This was attributed to the high capacity of the alloy for dislocation storage at various deformation temperatures. The serrated flow was observed at 773 and 873 K due to the dynamic strain aging (DSA) mechanism. Interestingly, in this case, the DSA mechanism has a beneficial effect on overall ductility, which was attributed to the increase in the strain hardening exponent result in a delay of necking. Texture examination also reveals the formation of a double fiber texture consists of a strong <111>//TA fiber and a weaker <001>//TA fiber due to the simultaneous contribution of twinning and octahedral slip {111} <110>.

---

[1] Corresponding authors.
E-mail address: zareih@ut.ac.ir (A. Zarei-Hanzaki)
E-mail address: habedi@iust.ac.ir (H. R. Abedi)
https://doi.org/10.1016/j.msea.2020.140600

# 1. Introduction

High entropy alloys (HEAs) have been recently introduced as one of the most attractive topics in alloying design strategies [1]. Exceptional effects such as high mixing entropy, severe lattice distortion, sluggish diffusion, and cocktail effects significantly contribute to their unique mechanical properties, excellent thermal stability, and outstanding wear and corrosion resistance [2-8]. The presence of a minimum five elements in equiatomic concentrations is generally considered to maximize the configuration entropy and production of a single-phase solid solution [9]. However, the Gibbs free energy is recently considered as the decisive factor to define the equilibrium state. This new approach leads to the design of single-phase HEAs consist of four or more than four elements, even in a non-equiatomic percentage [10,11].

Numerous scientific researches have been conducted up to now to investigate the deformation behavior of Cantor, the most well-known single-phase HEA by the nominal composition of CoCrMnFeNi holding FCC crystallographic lattice [12]. Despite considerable ductility and fracture toughness at cryogenic temperatures, this HEA possesses mechanical performance the same as conventional binary Fe–Mn alloys at room temperature. The underlying reason is the transition of the dominant deformation mechanism from deformation twinning at cryogenic temperatures to dislocation slip at room temperature [13]. The activation of dominant deformation mechanisms in HEAs is principally believed to be controlled by the stacking fault energy (SFE) [14]. It has been shown that tailoring the atomic ratios of individual components would reduce the SFE value and in turn the phase stability of HEAs at room temperature [15]. In this respect, metastable high entropy alloys (MSHEAs) have been designed [16] which benefit from assistant strengthening mechanisms like twinning induced plasticity (TWIP) or transformation induced plasticity (TRIP) effect in addition to the unique strengthening mechanisms in HEAs, such as stacking fault energy fluctuation, lattice distortion, and local concentration fluctuation [17-20]. For instance, Deng et al. [21] have designed a non-equiatomic single-phase MSHEA with a nominal composition of $Fe_{40}Mn_{40}Co_{10}Cr_{10}$ in which the TWIP effect leads to enhancement of room temperature mechanical properties in comparison with the well-known CoCrMnFeNi alloy.

In the case of Low SFE FCC alloys, dissociation of a perfect dislocation into a pair of Shockley partial dislocations can be described as a burger vector splitting: $1/2[110] \rightarrow 1/6[211] + 1/6[121]$. Such partial dislocations glide on every two consecutive {111} close-packed planes would generate deformation twins [22]. The twin boundaries act as obstacles to dislocation glide leading to enhancement of strength and ductility simultaneously [23]. The contribution of deformation twinning in work hardening depends upon the extent of the twinning activity. Therefore, the determining role of the chemical composition, grain size, grain orientation, strain rate, and deformation temperature should be considered [24]. Wu et al. [25] investigated the effect of grain size on the work hardening and twinning behavior of $Al_{0.1}CoCrFeNi$ alloy. The monotonically decreasing trend of the work hardening rate in fine-grain structures was attributed to the wider twin thickness and smaller twin spacing (i.e., lower twinning activity). Min et al. [26] also have discussed the relatively low critical strain of twinning and ease of twin nucleation in CoCrFeMnNi alloy in dynamic deformation. The competition of dislocation/ twin hardening and thermal softening and their contribution in work hardening under the various strain rate have been also discussed.

Coming to the point, it is obvious that in the case of TWIP HEAs, most previous reports have been devoted to room temperature deformation behavior. The present authors believe that the deformation temperature can significantly influence the work hardening behavior, strength/ductility balance, microstructure, and texture evolutions; and a more detailed and systematic study is highly necessitated. This would be overemphasized considering the fact that MSHEAs could be considered as promising candidates for the structural applications operating under varying sever conditions regarding temperature and applied stress. In this respect, the present work aims to study the deformation behavior of a MSHEA in tensile mode and in a wide range of temperatures between 298-873K. This has been supported in-details by characterization of the microstructure/ texture evolutions to set the microstructure-processing-properties relationship.

## 2. Experimental procedure

### 2.1. Alloy preparation

The investigated alloy in this study was a novel HEA by nominal composition $Co_{35}Cr_{20}Ni_{15}Fe_{15}Mn_{15}$ (at. %) designed by Wei et al. [27]. The alloy was prepared by vacuum arc melting using pure elements (purity higher than >99.9%), and subsequently, electro slag re-melting was used to avoid the segregation of the alloying elements. The exact chemical composition of the material is measured by XRF analysis, which is given in **Table 1**. The as-remelted material was homogenized at 1473 K for 5 h to improve the chemical homogeneity. The homogenized alloy was subsequently hot rolled at 1473 K by a 60% reduction in thickness to break the solidified structure. Finally, the alloy was homogenized at 1373 K for 2 h in argon atmosphere followed by water quenching.

### 2.2. Mechanical testing

The dog bone tensile specimens with gauge dimensions of 25×6×1.1 were cut from the as-homogenized material by electrical discharge machining (EDM) with the tensile axis aligned along the rolling direction. The uniaxial tensile tests were performed in the temperature range of 298-873 K under the strain rate of $0.001s^{-1}$ up to the fracture. For this purpose, a Gotech AI-7000 universal testing machine coupled with a programmable resistance furnace was utilized. Before the high temperature tensile tests, specimens were isothermally soaked in the specified deformation temperature for 7 min. In order to preserve the deformed microstructures, the fractured specimens were immediately quenched in water.

### 2.3. Microstructural Analysis

The specimens were cut near the fracture tip along the rolling direction and electron backscatter diffraction (EBSD) analysis was conducted using a field emission scanning electron microscope (FE-SEM, Quanta 3D FEG, FEI Company, USA) operating at 20 kV, equipped with a Hikari EBSD detector. The specimens were mechanically polished after grinding up to 1200 grit SiC paper followed by 1-3 μm water-based diamond suspension. Subsequently, a 50 nm colloidal silica solution as a final polishing followed by ultrasonic cleaning in acetone was implemented. The acquired data were collected using various step sizes (0.8μm, 0.15μm, and 0.05μm). The EBSD data were interpreted using orientation imaging microscopy (OIM) analysis software (TSL OIM analysis 7). The phase identification analysis by X-ray diffraction (XRD) was carried out using a Philips X'Pert PRO apparatus, using Cu-Kα radiation with a wavelength of 1.5406 Å operated at 40 kV and 30 mA between 40 and 100° (2θ).

## 3. Results and Discussion

### 3.1. Initial microstructure

The Inverse Pole Figure (IPF) map of the starting microstructure in the as-homogenized condition is shown in Fig. 1a is representing a recrystallized microstructure holding equiaxed grains. The average grain size, excluding twin boundaries (TBs), determined by a linear intercept method, is found to be 70 μm. The high fraction of annealing twins boundaries may indicate the low SFE of the experimented materials according to the principle of "ab initio" calculations [28,29]. The average grain size decreases to 41 µm by considering TBs as high angle boundaries (HABs). For more precise characterization, the boundary map of the initial microstructure has been depicted in **Fig. 1b**. The HABs, low angle boundaries (LABs: misorientation: 2-15º), and Σ3 TBs are depicted in black, red, and blue, respectively. Due to the previous homogenization process at elevated temperature, the formation of LABs is hindered, and their fraction is negligible. The corresponding misorientation angle distribution (**Fig. 1c**) exhibits a peak at about 60° which is related to the presence of a significant fraction (about 0.6) of Σ3 annealing twin boundaries. This is identified by 60°/<111> values in the misorientation angle and misorientation axis. The corresponding x-ray diffraction pattern (**Fig. 1d**) also confirms that the initial specimen contains a single phase with FCC lattice structure.

## 3.2. Mechanical properties and work hardening behavior

### 3.2.1. Strength/ductility balance

The typical stress-strain curves of the experimented alloy deformed at various temperatures are given **Fig. 2a**. The corresponding ultimate tensile strength (UTS), 0.2% offset yield strengths (YS), total elongation to fracture (TE), and uniform elongations (UE) are summarized in **Table 2**. As is realized, all of the strength and formability parameters possess their maximum values at room temperature (585 MPa, 227 MPa, 93%, and 86 %for UTS, YS, UE, and TE, respectively). **Fig. 2b** compares the UTS and TE of investigated alloy with those reported in previous researches for FCC HEAs, commercial, and advanced high strength steels [29]. Interestingly, the investigated alloys exhibit an extraordinary TE than that of other FCC alloys with comparable UTS. Another worth point which should be considered is the relatively low temperature sensitivity of the mechanical properties, in a way that the alloy exhibits an acceptable strength/ductility balance even at higher temperatures. By increasing the temperature upto 873 K, the UTS, YS, and TE decrease down to the 408MPa, 146MPa and 61%, respectively.

In a more detailed view, the temperature dependency of YS can be assessed by the consideration of various strengthening mechanisms which contribute as follows:

$$\sigma_{YS}(T) = \bar{\sigma}_{Fr} + \Delta\sigma_{Gr} + \Delta\sigma_{Dis} + \Delta\sigma_{Pr} \tag{1}$$

where $\bar{\sigma}_{Fr}, \Delta\sigma_{Gr}, \Delta\sigma_{Dis}, \Delta\sigma_{Pr}$ are the contributions from average lattice friction, the grain boundary strengthening, initial dislocation strengthening, and precipitation strengthening, respectively. The $\bar{\sigma}_{Fr}$ contribute to the thermal part, and the remaining ones mostly contribute to the athermal part of YS [30]. According to Ref [31], $\Delta\sigma_{Gr}$ and $\Delta\sigma_{Dis}$ are approximated to be zero due to the relatively initial coarse grains and the well-annealed condition of starting microstructure. Besides, $\Delta\sigma_{Pr}$ can be eliminated, considering the single-phase microstructure of the alloy at various conditions (refer to the XRD diffraction patterns in **Fig. 3a**). Having all into consideration the absence of any possible microstructural barrier against the dislocation motions, the temperature dependency of YS in the present case is believed to be caused only by the average lattice friction ($\bar{\sigma}_{Fr}$). This is considered a combination of solid solution strengthening and intrinsic lattice friction, which stems from intrinsic heavily lattice distortion and particular dislocation configuration [32,33]. In opposition to conventional FCC metals and alloys, the atomic arrangements are continuously varying along the dislocation line in FCC HEAs. Hence, the dislocation line energy is varying in relatively wide values [33]. This will affect the dislocation mobility during the overcoming the lattice friction stress is represented by the Peierls stress, which is given by:

$$\sigma_P = \frac{2G}{1-\nu} \exp\left(\frac{-2\pi\omega}{b}\right) \tag{2}$$

where $G$ is the shear modulus, $\nu$ is the Poisson's ratio, $\omega$ is the dislocation width, and $b$ is the Burgers vector [30]. As it was mentioned former, HEAs exhibit local variations in $\omega$ and $b$ values. Therefore, it will lead to continual fluctuation Peierls barrier height as a function of temperature, which also affects the dislocation mobility [34]. Nevertheless, as shown in **Fig. 3b**, the experimented alloy shows a lower decreasing rate of YS vs. temperature than those found for the well-known CoCrFeMnNi HEA holding FCC structure and comparable grain size [35,36]. The authors believed that further investigations would be needed to interpret this behavior that is beyond the scope of present work and will be discussed in our future works.

As is seen in **Fig. 4a**, the investigated alloy shows an extraordinary TE at a wide range of deformation temperatures compared with other FCC HEAs and CAs [37]. The overall ductility of the experimented alloy at various deformation temperatures can be estimated by strain hardening exponent (n). It is calculated by true stress- true strain curves fitted with Ludwick equation:

$$\sigma = \sigma_{YS} + K\varepsilon_p^n \tag{3}$$

where $K$, $\sigma_{YS}$, $\sigma$, and $\varepsilon_p$ are the strength coefficient, yield stress, true stress, and true plastic strain, respectively [38]. The n and $K$ results for various deformation temperatures obtained from equation 3 are listed in **Table 2**. A high n value (almost equal to 1) of the investigated alloy at room temperature is unique as compared with the conventional metals and alloys [39-41]. Furthermore, it remains large at a wide range of deformation temperature, which indicates the high work hardenability and high dislocation storage capacity, which increased the ductility by delaying the onset of necking.

Demonstration of such significant mechanical properties are directly related to the work-hardening rate ($\Theta$) of the alloy during straining. **Fig. 4b** exhibits the true stress-strain curves superimposed with their corresponding Θ plots according to the Considère criterion. (in order to plot the Θ curves, true stress-strain curves were fitted by the nine order polynomial curves). As depicted in **Fig. 4b**, the specimen deformed at 298 K, has a three-stage Θ that continuously decreases in the first stage (2%<ε<10%). It is related to the transition of the elastic to plastic deformation behavior accompanied by a competition between the storage and the recovery, which is typical for the low SFE alloys [42]. Generally, the FCC HEAs, such as CoCrFeMnNi, reveal a monotonous decay of the Θ by further straining [43]. However, the present MSHEA exhibits a relatively persistent Θ in stage II (10%<ε<57%). In stage III (57%<ε<$ε_{fracture}$=65%), the Θ decrease monotonically upon necking and fracture.

Two considerable features in the work hardening behavior of investigated alloy upon deformation at various temperatures can be mentioned. Firstly, the maximum magnitude of Θ decreases less than 400 MP by increasing the deformation temperature from 298K to 873K. The mentioned observation indicates a relatively-stable competitive relationship between strain hardening and strain softening at various deformation temperatures that will be discussed in more detail in the next section. Secondly, as can be observed, the Θ of the specimens deformed at 773K and 873K exhibit a slight increase in the length of stage II, which shows a deviation rather than the previous trend. Moreover, at the higher strain levels, the magnitude of Θ for specimens deformed at 773 and 873K is higher than 573 and 673 K upon to necking. Change in the work hardening behavior in the experimented alloy is accompanied by serration in flow behavior.

### 3.2.2. Dynamic strain aging

The serrated flow behavior is clearly identified at 773 K and 873 K, the extent of which is influenced by the amount of strain and deformation temperature. The serration was initiated after yielding point at 873 K, but for 773 K appeared after the critical engineering strain of about 0.34.

These can be considered as a strong clue for the occurrence of dynamic strain aging in specified temperature range. Serration behavior in the flow curves of conventional alloys (CAs) has been reported in numerous researches attributed to dynamic strain aging (DSA) phenomena. Specifically, it is believed that the pinning and unpinning of dislocation by the interaction of substitutional or interstitial solute atoms with dislocation lines is the origin of DSA. It leads to instability in deformation and promotes the formation of Portevin–Le Chatelier (PLC)bands on the deformed specimens [40].

The serrated flow behavior has been reported in a few studies in HEAs. The coarse-grained mean-field theory (MFT) model interpreted the serration behavior of some medium and high entropy alloys as avalanches of slipping weak spots along to slip planes and re-strengthening of them during the time intervals between slip avalanches [44]. Wang et al. [45] used the atomic and electronic basis for interpreting the serration behavior of some refractory HEAs. They proposed that the presence of strongly bonded clusters and weakly bonded glue atoms result in avalanches of defect movement and cause instability in macroscopic flow behavior. Fu et al. [36] derived two activation energies for serrated flow behavior of CoCrFeMnNi HEA over different temperature ranges and concluded that the solute pinning of dislocations controlled by pipe diffusion and cooperative lattice diffusion of constituent atoms in different temperature. Recently, Tsai et al. [46] successfully suggested a theoretical calculations-based mechanism for the interaction of constituent atoms of a HEA with dislocations result in PLC band formation and serration flow behavior. They stated that the pining of dislocations is happened by in-situ rearrangement of substitutional solute atoms around the cores of dislocations. Despite classic long-range diffusion theories, this new mechanism requires a very short time local diffusion for solute atoms to catch moving or impeded dislocations.

In this respect, it is assessed that for the specimen deformed at 773K, the solute mobility around the core of dislocations is limited. Hence, the locking/unlocking events do not take place in the lower strains. By further straining, dislocation density increases result in ease of the aging of dislocations. However, the frequency of the serrations on the stress-strain curve is relatively low. On the other hand, in the case of higher deformation temperature (873K), the solute mobility is enhanced, promoting the instantaneous locking of dislocations by the short time local diffusion. Hence, a high frequency of serrations can be observed. Type of serration at 773K and 873K is referred to as type A and B, respectively [40]. For a more precise assessment of the mechanical properties, microstructure evolutions of the experimented alloy were characterized as a function of temperature in the next section.

### 3.3. Microstructure evolutions upon deformation

### 3.3.1. TWIP effect

Fig. 5 shows inverse pole figure (IPF) maps of the microstructure of the specimens which have been deformed up to fracture at different temperatures. The reference axis corresponds to the tensile axis, and the reference colored triangle is also given. As is evident, the fraction of annealing twins shows a strong decline compared with the initial microstructure (Fig. 1a). It can be attributed to progressively decaying of the twin/matrix misorientation relationship due to increasing local lattice rotations needed to maintain the strain compatibility with increasing tensile strain [47]. However, there are parallel or intersecting bands with various thicknesses and spacings within the elongated grains. In order to more detailed characterization, scanning with a lower step size (50 nm) was conducted, the result of which is given in Fig. 6a. The point to point and point to origin misorientation profiles along with the line A shows a 60 ° misorientation angle, indicating that these bands can be considered as deformation twins (Fig. 6b). In addition, the local pole figure of

{111} plane indicates the one-point coincident pattern (Fig. 6c), which further verify the twin relationship of these specified bands with around the FCC matrix. The formation of deformation twins may occur at the early stage of deformation, and they start to thicken and form twin bundles where the level of plastic deformation is increased. The generated twins from different systems will interact with each other and make a dense twined substructure. Also, the interaction between partial dislocations and twin boundaries may cause deviation of Σ3 twin boundaries from the ideal misorientation angle or disorientation axis; i.e. deviation from the ideal coincidence of 60°<111>.

Kernel average misorientation (KAM) exhibits the local strain distribution and could qualitatively evaluate the contribution of the twinning mechanism in the work hardening rate [48]. Fig. 6d illustrates the Kernel average misorientation (KAM) of the microstructure, which has been deformed at 298K. As is depicted by black circles, the grain partitioning is evident, and the effective grain size decreases due to progressive fragmentation of the grains by deformation twins. Each twin-surrounded area will act as a grain; therefore, the mean free pass of dislocations will be reduced during the formation of deformation twins. This would increase the capability of work hardening through the "dynamic Hall-Petch effect" [49]. In addition, as is indicated by the black arrows in Fig. 6d, higher values of KAM were observed near and within the twin bundles and twin-twin intersections. The possible reason for this observation is the dislocation-twin and twin-twin interactions. These interactions caused accumulation of sessile dislocation near and within the twin boundaries, which hinder dislocation motions and provide more dislocation multiplication result in enhancement of $\Theta$ value [50].

Based on the mentioned observations, the noticeable plateau in stage II of $\Theta$ (10%<ε<57%) can be attributed to the contribution of deformation twinning in work hardening as a source of dislocation multiplication. By progressive fragmentation of grains, critical resolved shear stress for the twin nucleation enhances. The Low nucleation rate of deformation twins accompanied by saturation of microstructure by high dislocation density results in reducing dislocation multiplication [43]. Therefore, the overall magnitude of Θ decreases in higher strains and stress localization leads to necking and failure.

Fig. 5 also elucidates that grains tend to orientate mainly with the <111> or <001> crystallographic direction parallel to the tensile axis (TA) direction. Twinning was apt to appear in some grains rather than others. Dependence of deformation twinning on grain orientation can be determined by Schmid factor analysis [51]. The orientations of grains were close to the < 111 > // TA, which have larger Schmidt factor for the twinning (0.31) than that for perfect dislocation slip (0.27) are favor for twinning than other orientations. While in the orientations of grains close to the <001> // TA, perfect dislocation slip is predominant because of a larger Schmidt factor for dislocation slip (0.41) than that for deformation twinning (0.24). Correlation between the grain orientation and dominant deformation mechanisms also is reported in the previous works [21]. Furthermore, it is reported that the width of stacking faults as precursors of deformation twins could show an orientation dependence. In <111> oriented grains, the Shockley partial dislocations dissociated completely and ease the production of stacking faults, but in <001> oriented grains, the partial dislocations will experience constriction and intend to combine instead of dissociation [52].

Another feature is that, with increasing deformation temperature, the fraction of deformation twins decreased, and the secondary twins are not formed at higher temperatures. In this regard, Fig. 7a-d illustrates Image Quality (IQ) superimposed by Σ3 twin boundary maps for the fractured specimens at different temperatures. The blue color indicates the Σ3 twin boundaries. As can be observed at 673 K, Σ3 twin boundaries still are detected in a few grains while the fraction of them dropped drastically. For all temperatures above 673 K, deformation twinning is suppressed completely and twin boundaries captured above 673 are the initial annealing twin boundaries. Albeit, EBSD is merely used to show the trends, and due to the limitation in spatial resolution of EBSD, thinner deformation twins may not be detected completely.

Deformation temperature affects the twinning activities in the following ways. (i)The SFE, which controls the activation of the dominant deformation mechanism, is a strong function of temperature. As deformation temperature increased from 298K to 873 K, SFE increases and leads to reducing the width of stacking faults. It means enhancing critical stress for nucleation of deformation twins, result in the suppression of twining nucleation. (ii) Moreover, increase the deformation temperature triggers the cross slip, and the annihilation of dislocations results in suppresses the formation of stacking faults as nuclei of deformation twins.

### 3.3.2. Boundaries evolutions

In addition to Σ3 twin boundaries, deformation temperature has an undeniable effect on random boundaries, which requires detailed characterization of microstructure. **Fig 8a-d** illustrates the boundary maps of the specimens after fracturing at different temperatures. Low (misorientation: 2-5º and 5-15º) and high angle boundaries (misorientation>15º- except Σ3 twin boundaries) delineated in red, green, and black, respectively. For quantified characterizations of boundaries, misorientation angle distributions are presented in **Fig 9a-d**. In the deformed microstructure, there are HABs with a range of about 60º that belong to the Σ3 twin boundaries. From **Fig 9a,** it can be deduced that in comparison with the initial specimen (**Fig. 1c**), the fraction of the LABs has been increased, and consequently, the fraction of the HABs has been decreased by deforming the investigated alloy. The significant rise in the fraction of LABs can be rationalized in terms of dislocation accumulation during plastic deformation. The low SFE of investigated alloy leads to some specific dissociations and interactions of dislocations that promote the formation of Lomer and Lomer-Cottrell (L-C) dislocation locks. These sessile dislocation locks act as dislocation barriers and sources on non-primary slip planes and contribute as an additional source of dislocation multiplication [53]. Dislocation activity, including slip, climb, and cross slip, is thermally activated processes and will be intensified as temperature increases [54]. Therefore, by increasing deformation temperature up to 873K, the annihilation of dislocations enhanced by dislocation motions and leads to a gradual reduction in LABs fraction (**Fig 9b-d**).

For a complete analysis of processed microstructures, the KAM of deformed specimens is depicted in **Fig. 10**. The higher KAM values are obvious for the deformed specimen, which can be attributed to the dislocation generation. As expected, twin and grain boundaries contain higher KAM value than those in the matrix, especially in heavily deformed regions (red color) due to higher dislocation accumulation in these regions. The sample deformed at room temperature shows higher KAM values compared to those deformed at elevated temperatures. This may make sense since, at higher deformation temperatures, dislocation annihilation occurred.

From the correlation of microstructural evolutions and work hardening behavior, it was deduced that the persistent work hardening behavior at room temperature is due to the low SFE, which promotes planar slip and twinning activity. It leads to increased dislocation storage and accumulation, which together with the freshly presented twin boundaries increase the overall magnitude of $\Theta$. As a result, the onset of necking shifted to higher strains, and subsequently, more pronounced strength and ductility are achieved. The magnitude of $\Theta$ is the result of two competitive factors; The dislocation generation rate and the dislocation annihilation rate. The annihilation rate significantly affected by temperature; therefore, an increment in deformation temperature up to 673K eases the dislocations annihilation and suppresses twining activity. Hence, the dislocation-dislocation and dislocation-twin interactions are reduced, which will reduce the overall magnitude of Θ. The specimens deformed at 773 and 873K (DSA regime) show an unexpected behavior in terms of the Θ and ductility variations. Generally, the DSA mechanism causes instability in plastic deformation and promotes deformation localization, linked with a loss in ductility [40]. However, in low SFE alloys, increasing ductility at the DSA regime has been attributed to the formation of deformation twinning [55]. Nevertheless, no deformation twinning is observed in the investigated alloy. The beneficial effect of the DSA mechanism in the experimented alloys was believed to favor the planar dislocation slip [56]. This manifests in increases the magnitude of Θ and n value at 773K and 873K, resulting in a delay of necking and prolongs the overall ductility.

### 3.4. Texture evolutions upon deformation

The texture evolutions of the investigated alloy have been examined in terms of EBSD-derived pole figures (PF) and orientation distribution functions (ODF). **Fig. 11** shows the {001} and {111} pole figures of the initial and deformed specimens at different temperatures. A weak texture is characterized in the case of the homogenized condition by the maximum intensity of 2.421, which is typically considered as annealing texture characteristics of FCC alloys (**Fig. 11a**) [57]. Where the investigated alloy is tensile deformed at 298K, the reduction in the intensity of some texture components is observed (**Fig. 11b**). The texture evolution during deformation is discussed in connection with the activity of slip and twinning systems. Due to the activity of {111}<110> slip systems, unstable crystal orientations gradually rotated toward the stable <111> and <001> orientations holding a high and a low Taylor factor, respectively [58]. Such rotations result in the development of a double fiber texture parallel to TA. This double fiber consists of a strong <111>//TA fiber with an intensity of 9.528 and a weaker <001>//TA fiber with an intensity of 3.549. As is seen in **Fig. 11,** the maximum intensity of <001>//TA fiber is found at 298K. The reason for the higher intensity of <001> fiber at 298K is attributed to a higher fraction of deformation twinning. New orientations are formed where twins form inside the <111 >//TA oriented grains, thus strengthen the <001> fiber [59]. Additionally, high intensity of <001> fiber before the deformation may effectively contribute in ultimate intensity of this fiber. By increasing deformation temperature up to 873K, due to the more activity of different slip systems the overall intensity of double fiber is reduced and orientations are dispersed gradually (**Fig. 11c-e**).

The ideal texture components of FCC materials are shown schematically for $\varphi_2 = 0°$, 45°, and 65° ODF sections in **Fig. 12** and **Table 3** [60]. In the present case, the initial texture comprises of cube (C) components as the main component and a relatively weak rotated cube (Rt-C) and brass (Bs) in the $\varphi_2 = 0°$ (**Fig. 13a**). By examining the texture in $\varphi_2 = 45°$, next to the observed components of texture in the previous section, Copper (Cu) and Copper Twin (Cu(T)) are also observed with relatively weak intensity. In $\varphi_2 = 65°$, only C component is observed. The mentioned components by the maximum intensity of 7.3 are carried over from the applied thermomechanical and homogenization process before the tensile test. Some components cannot be detected, indicating that there is a random texture as well.

Through deformation at 298K, a pronounced increase in the intensity of the Rt-Cu and Bs (belonging to the <111> fiber) up to a maximum intensity of 12.523 is observed (**Fig 13b**). The increase in the intensity of these components is caused by specific grain rotations related to the slip activity. The higher intensity of the main texture components belonging to the <111> fiber promotes the twinning activity. Initial twinning activity is related to the presence of Bs orientation that is favorable for twinning [59]. Grains with this orientation possess higher twinning Schmid's factor than slip; see the Schmid factor values in **Table 3** [60]. Although the Cu component in the initial texture is favorable for twining in terms of Schmid's factor but the reinforcement of this orientation densities is not favored by deformation and consequently disappears in developed final texture [59].

Moreover, deformation twinning plays a key role in the texture development in FCC alloys. During straining, grains with the Cu-{112}<111> orientation first twin into the copper-twin (Cu(T))-{552}<115> orientation. Dislocation glide then causes the Cu(T)- {552} <115>-oriented grains to rotate towards the orientations that are close to the <001> fiber orientation. In grains with this orientation, dislocation slip is favored relative to twinning as the Schmid factors for dislocation slip is lower than that of twinning [61]. Grains with <001>//TA orientation exhibit the smallest rotation amongst all other orientations and have no preferred rotation direction. As is depicted in $\varphi_2 =0°$ and $\varphi_2 =45°$, this is associated with a slight increase in Rt-G component and a slight decrease in C component and concurrent disappearing of the Rt-C and Cu components. It should be noted that there is a high-intensity texture component in the $\varphi_2 =65°$ section. By careful investigation of the texture development in other sections, it was determined that this component is Bs.

The type of main texture components does not change by increasing deformation temperature since both octahedral slip {111} <110> and mechanical twinning are active up to 473K. While a decrease in the texture intensity of the Rt-Cu and Bs along with the <111> fiber, the Rt-G orientations strengthen to a slight extent (**Fig. 13c**). As deformation temperature rose to 673K next to the texture development in the previous temperature, the C component is eliminated completely (**Fig. 13d)**. At the 873K, twining activity diminished, and activation of different slip systems weaken the main texture components. In this regard, the intensity of Bs and Rt-Cu components are being decreased. Also, while the Rt-G component is completely disappeared, the C component emerged in a rather weak intensity (**Fig. 13e**).

## 4. Conclusion

In this investigation, the effect of deformation temperature on tensile behavior of $Co_{35}Cr_{20}Ni_{15}Fe_{15}Mn_{15}$ (at. %) HEA has been systematically studied based on the evolved microstructure and texture at the temperature range of 298-873K. The important findings can be summarized as follows:

- The mechanical twinning and planar slip were the dominant strain accommodation mechanisms of the investigated alloy which led to the significant microstructure refinement and complex twin/dislocation interactions. These caused a higher rate of dislocation accumulation and dense substructure development, where the low angle boundaries usually formed near the original grain boundaries and twin bands. Consequently, constant strain hardening regions were observed and extraordinary strain hardening exponents were achieved, which resulted in a strength/ductility balance at various deformation temperatures.

- Upon increasing the deformation temperature up to 673K, despite the suppression of twinning and intensification of the recovery process, the temperature sensitivity of the mechanical properties was relatively lower compared with other high entropy and conventional FCC alloys. This phenomenal finding was attributed to the excellent capacity of dislocation accumulation and storage. Interestingly, the occurrence of dynamic strain aging at elevated temperatures (773 and 873K) improved the hardenability by increasing both the strain hardening exponent and the work hardening rate values.

- In connection with the activity of specified slip and twinning systems, a sharp double fiber texture, with a strong <1 1 1>//TA fiber and a weak <0 0 1>//TA fiber, was produced. Initial mechanical twinning occurred due to the presence of Bs and Cu orientations of the weak initial texture and developed in grains having <1 1 1>//TA orientations. The double fiber texture was weakened through increasing the deformation temperature where the variation of specified texture components was obvious.

**Data Availability**

The raw/processed data required to reproduce these findings cannot be shared at this time as the data also forms part of an ongoing study.

Energies of NiFeCrCoMn High-Entropy Alloy, JOM. 65 (2013) 1780–1789. https://doi.org/10.1007/s11837-013-0771-4.

[16] S. Wei, F. He, C.C. Tasan, Metastability in high-entropy alloys: A review, J. Mater. Res. 33 (2018) 2924–2937. https://doi.org/10.1557/jmr.2018.306.

[17] C. Varvenne, A. Luque, W.A. Curtin, Theory of strengthening in fcc high entropy alloys, Acta Mater. 118 (2016) 164-176. https://doi.org/10.1016/j.actamat.2016.07.040.

[18] S.M. Vakili, A. Zarei-Hanzaki, A.S. Anoushe, H.R. Abedi, M.H. Mohammad-Ebrahimi, M. Jaskari, S.S. Sohn, D. Ponge, L.P. Karjalainen, Reversible dislocation movement, martensitic transformation and nano-twinning during elastic cyclic loading of a metastable high entropy alloy, Acta Mater. 185 (2020) 474–492. https://doi.org/10.1016/j.actamat.2019.12.040.

[19] Y. Zeng, X. Cai, M. Koslowski, Effects of the stacking fault energy fluctuations on the strengthening of alloys, Acta Mater. 164 (2019) 1–11. https://doi.org/10.1016/j.actamat.2018.09.066.

[20] S.S. Sohn, A. Kwiatkowski da Silva, Y. Ikeda, F. Körmann, W. Lu, W.S. Choi, B. Gault, D. Ponge, J. Neugebauer, D. Raabe, Ultrastrong Medium-Entropy Single-Phase Alloys Designed via Severe Lattice Distortion, Adv. Mater. 31 (2019) 1–8. https://doi.org/10.1002/adma.201807142.

[21] Y. Deng, C.C. Tasan, K.G. Pradeep, H. Springer, A. Kostka, D. Raabe, Design of a twinning-induced plasticity high entropy alloy, Acta Mater. 94 (2015) 124–133. https://doi.org/10.1016/j.actamat.2015.04.014.

[22] M. Ghiasabadi Farahani, A. Zarei-Hanzaki, H.R. Abedi, J.H. Kim, M. Jaskari, P. Sahu, L.P. Karjalainen, On the activation of alternated stacking fault pair twinning mechanism in a very large-grained Fe–29Mn–2.4Al steel, Scr. Mater. 178 (2020) 301–306. https://doi.org/10.1016/j.scriptamat.2019.11.035.

[23] V. Shterner, I.B. Timokhina, H. Beladi, On the work-hardening behaviour of a high manganese TWIP steel at different deformation temperatures, Mater. Sci. Eng. A. 669 (2016) 437–446. https://doi.org/10.1016/j.msea.2016.05.104.

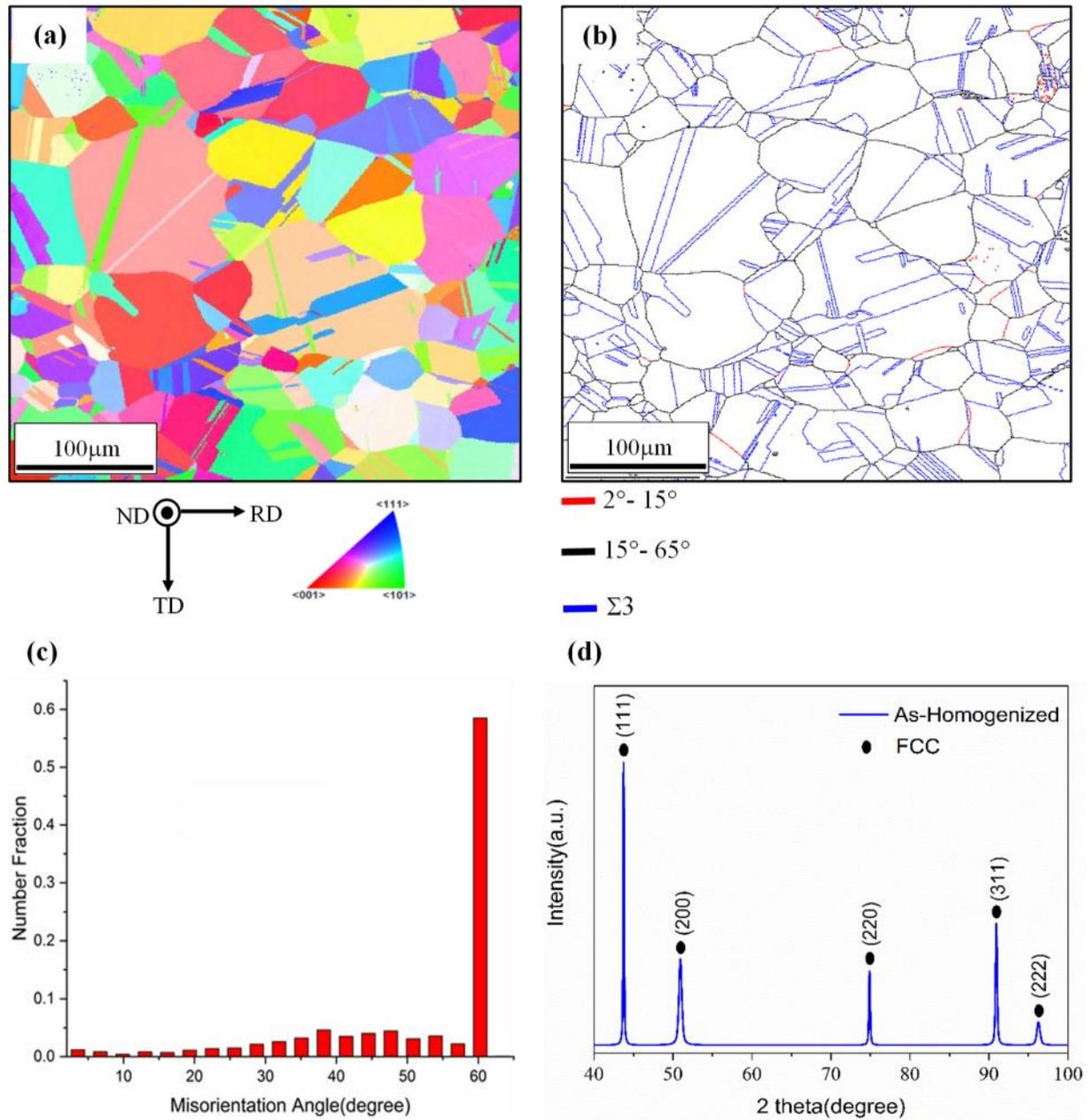

Fig. 1. Microstructure and phase analysis of initial specimen (a) EBSD IPF map showing fully recrystallized microstructure. (b) Grain boundary map (black, red and blue lines represent HABs, LABs and Σ3 TBs respectively. (c) Misorientation angle distribution. (d)X-Ray diffraction pattern showing the single FCC phase.

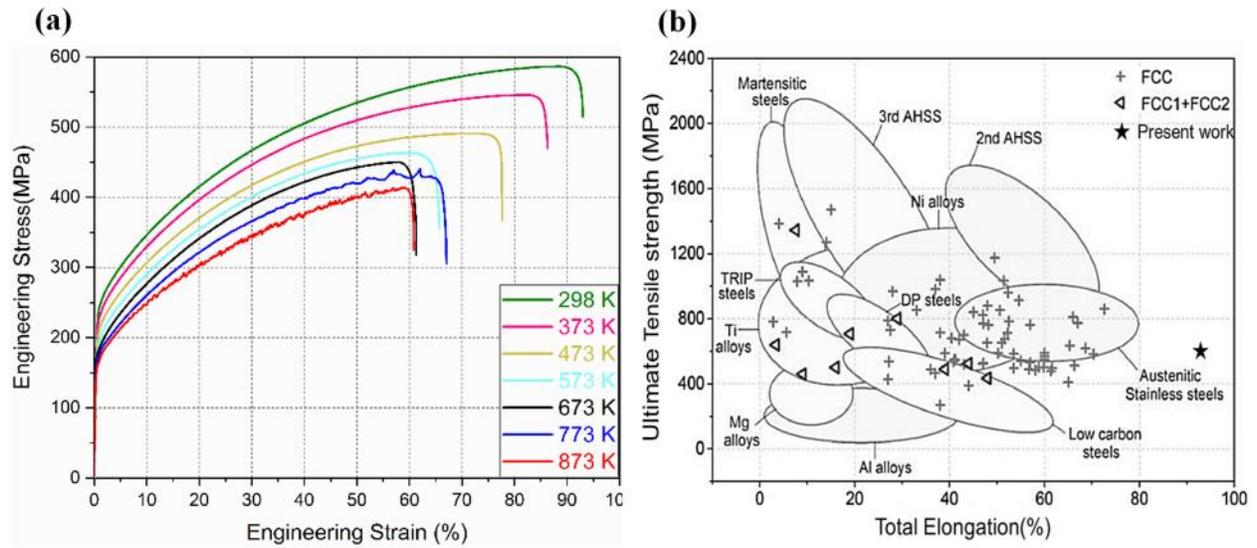

*Fig. 2. (a) Engineering stress-strain curves of the specimens deformed at 298-873 K. (b) Compares the uniaxial tension test data of investigated HEA with other FCC HEAs, selected commercial and advanced alloys. Adapted from [29].*

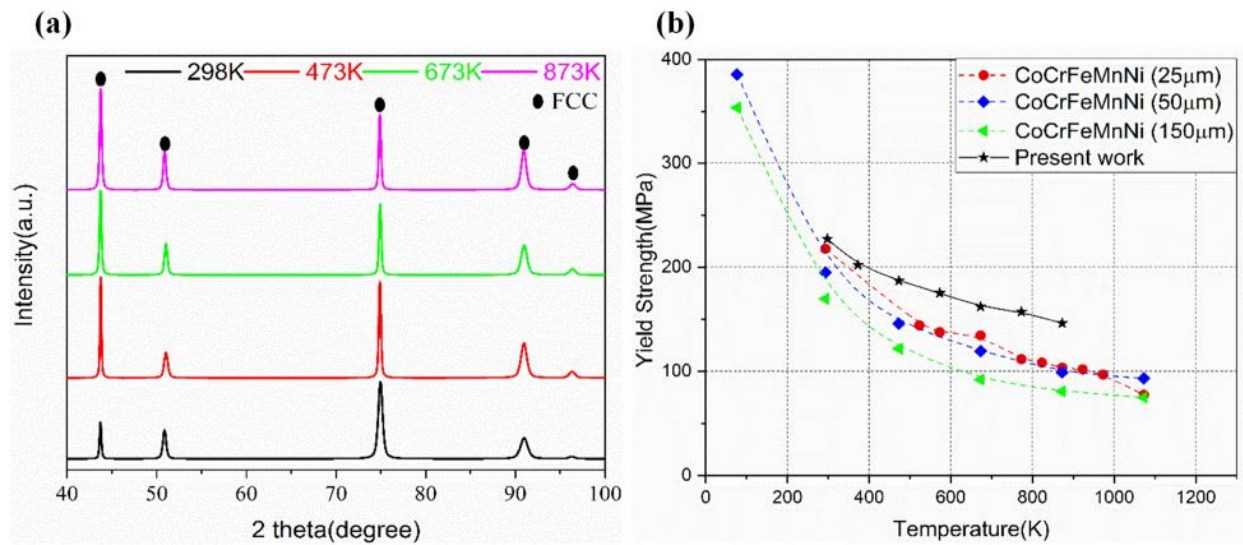

*Fig. 3. (a) XRD pattern of specimens after tensile tests at 298-873 K, (b) Temperature dependence of the Yield strength of investigated HEA($Co_{35}Cr_{20}Fe_{15}Mn_{15}Ni_{15}$) compare with $Co_{20}Cr_{20}Fe_{20}Mn_{20}Ni_{20}$. Adapted from [35,36].*

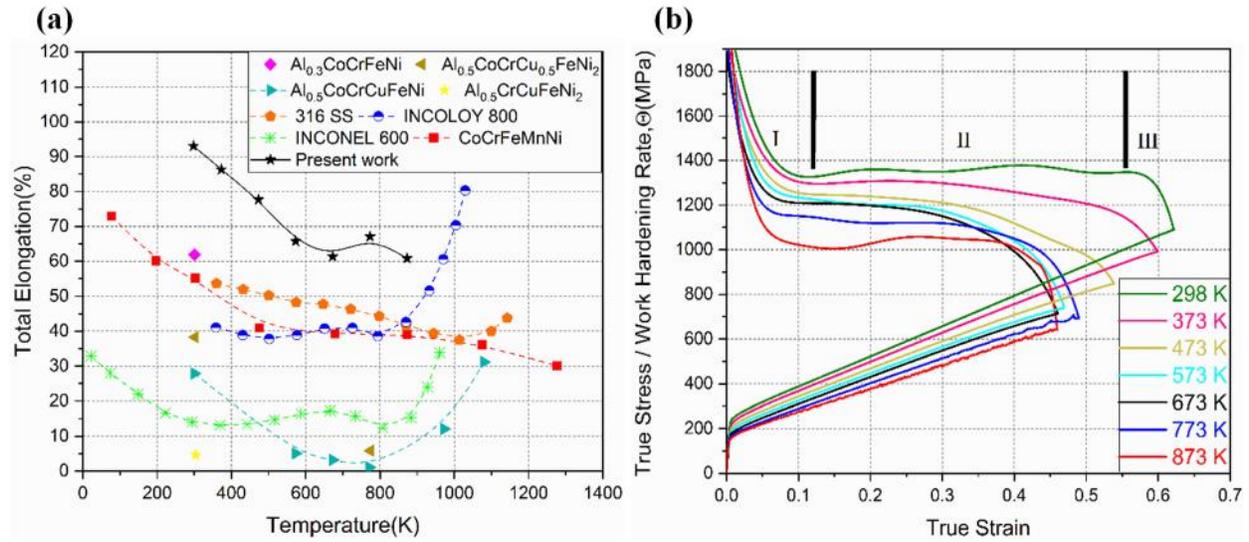

*Fig. 4. (a) Temperature dependence of the Total Elongation of investigated HEA(Co$_{35}$Cr$_{20}$Fe$_{15}$Mn$_{15}$Ni$_{15}$) compare with FCC base HEAs, selected commercial and advanced alloys. Adapted from [37], (b) True stress-strain curves superimposed with the corresponding ϴ curves of specimens deformed at 298-873K.*

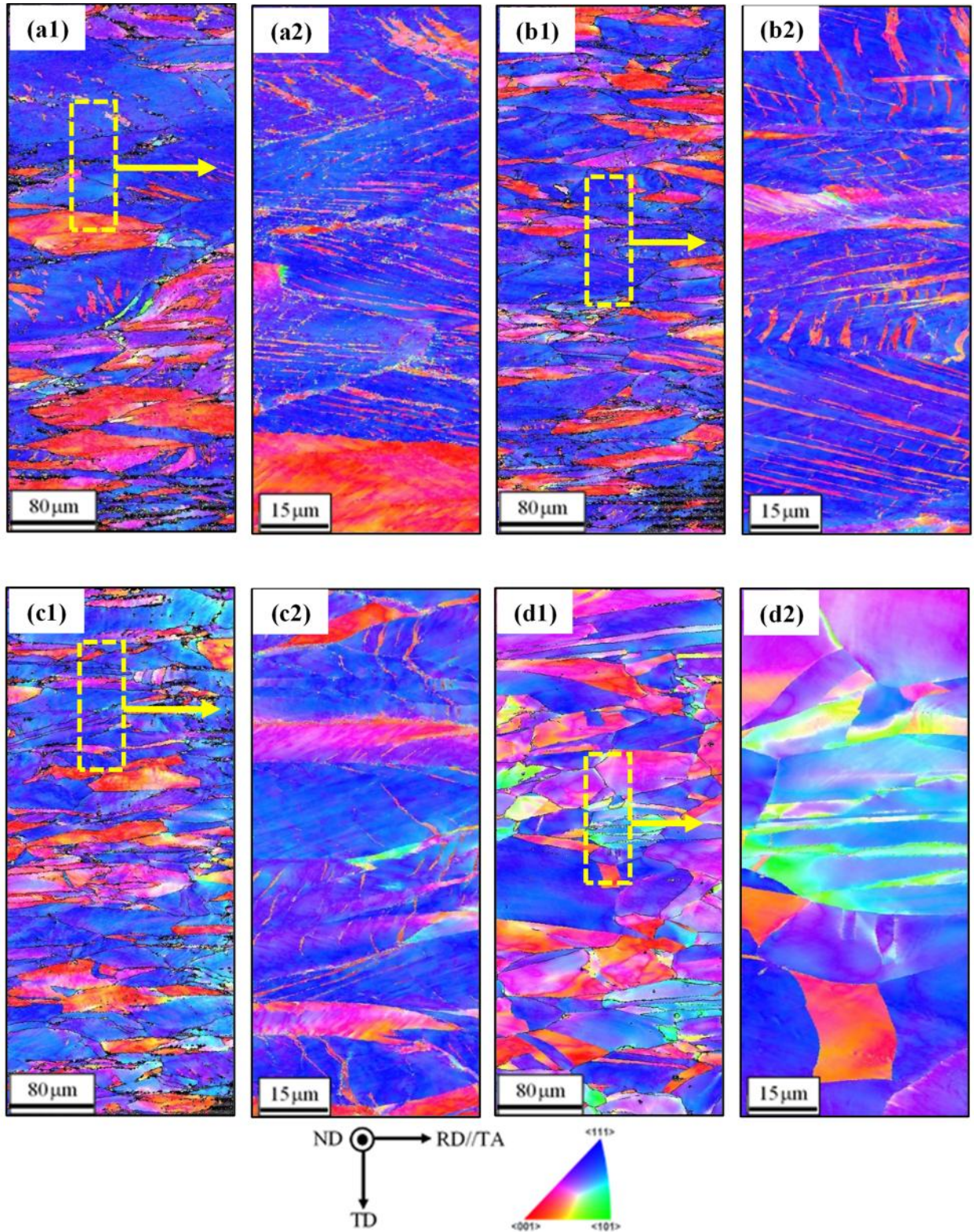

*Fig. 5. IPF maps of fractured specimens at temperature of (a1&a2)298 K, (b1&b2)473 K, (c1&c2)673 K and (d1&d2)873 K.*

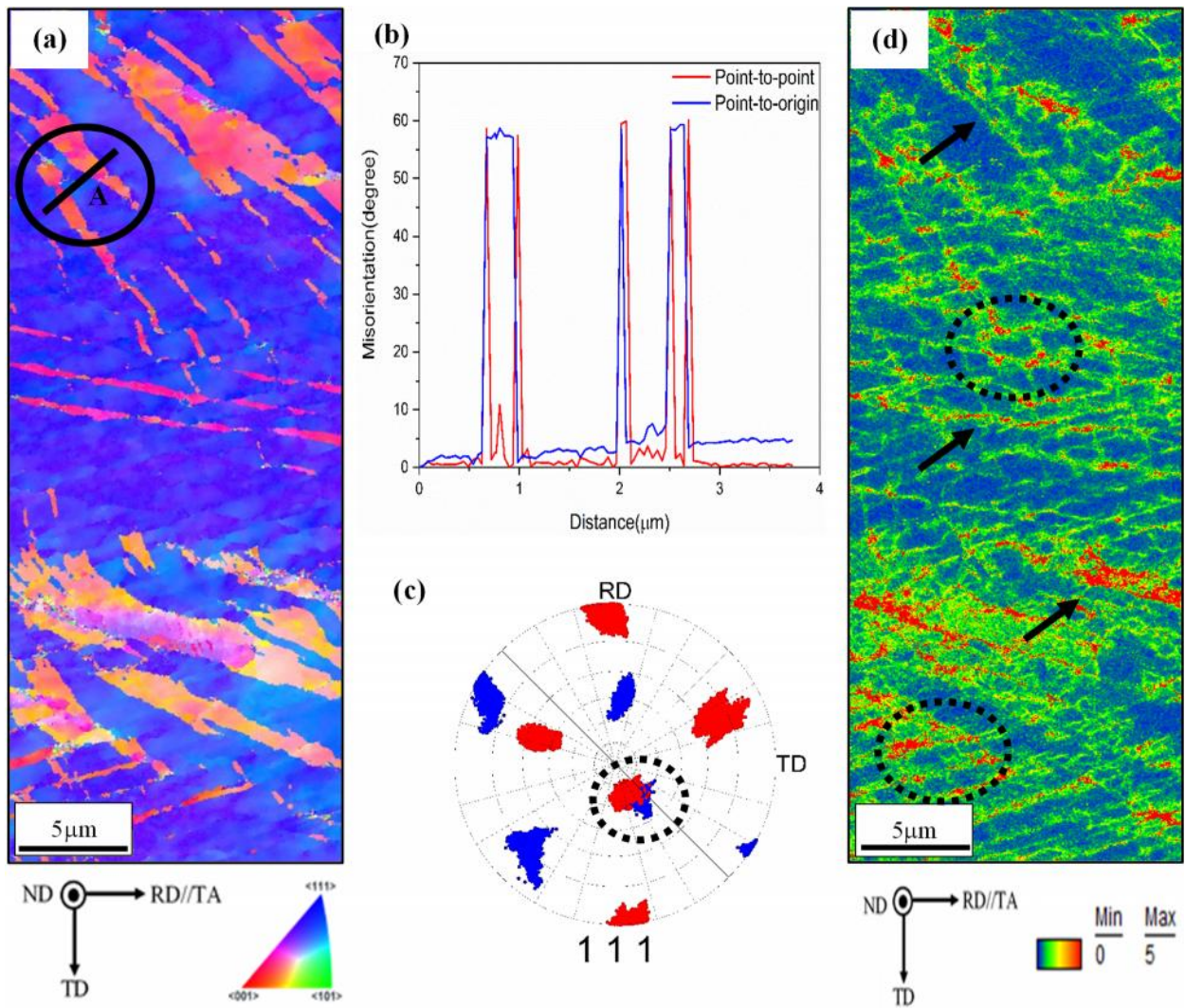

*Fig. 6. (a) IPF map of specimen deformed at 298K. (b) The point to point and point to origin misorientation profiles along line A in IPF map. (c) The local pole figure of {111} planes around twin band and adjacent matrix (black circle in IPF map). In corresponding pole figure the red and blue spots are attributed to the matrix and twin, respectively. (d) The kernel average misorientation (KAM) of fractured specimen at 298K. In corresponding KAM black arrows exhibit high KAM value near and within the twin bands and black circles surround twin substructures.*

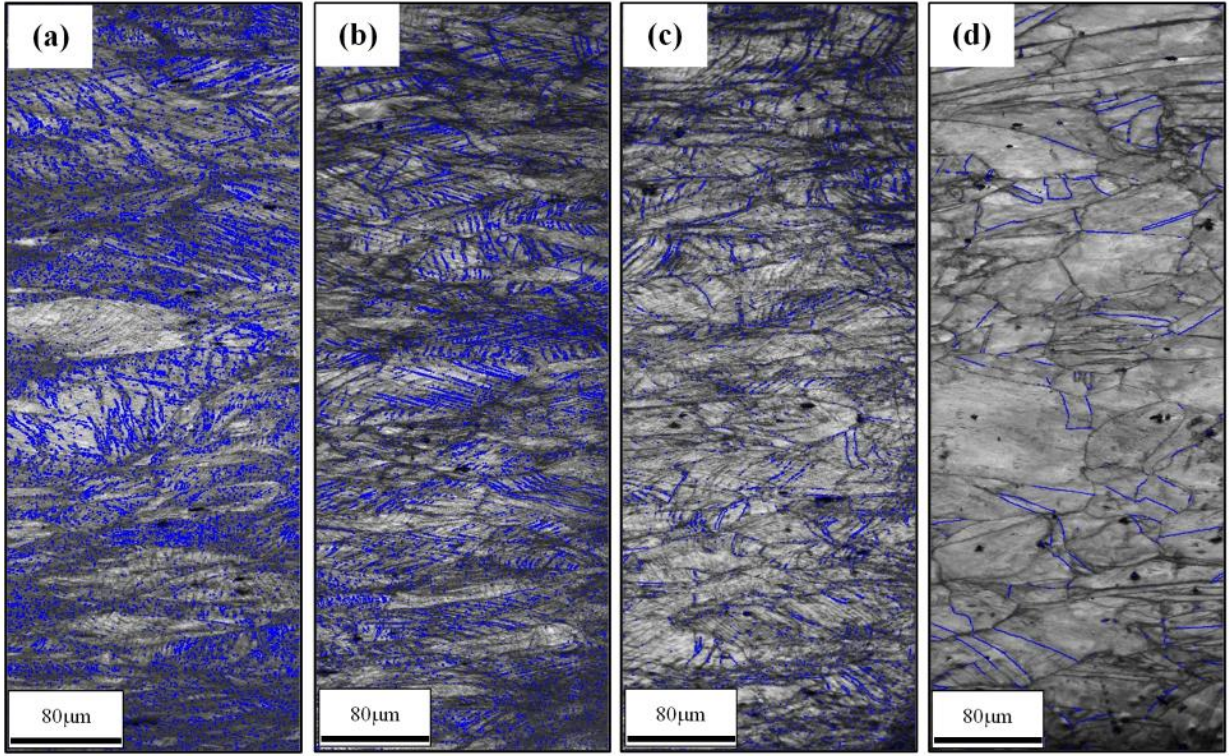

Fig. 7. The IQ map of fractured specimens at (a) 298 K, (b) 473 K, (c) 673 K, and (d) 873 K. The Σ3 twin boundaries are delineated in blue color.

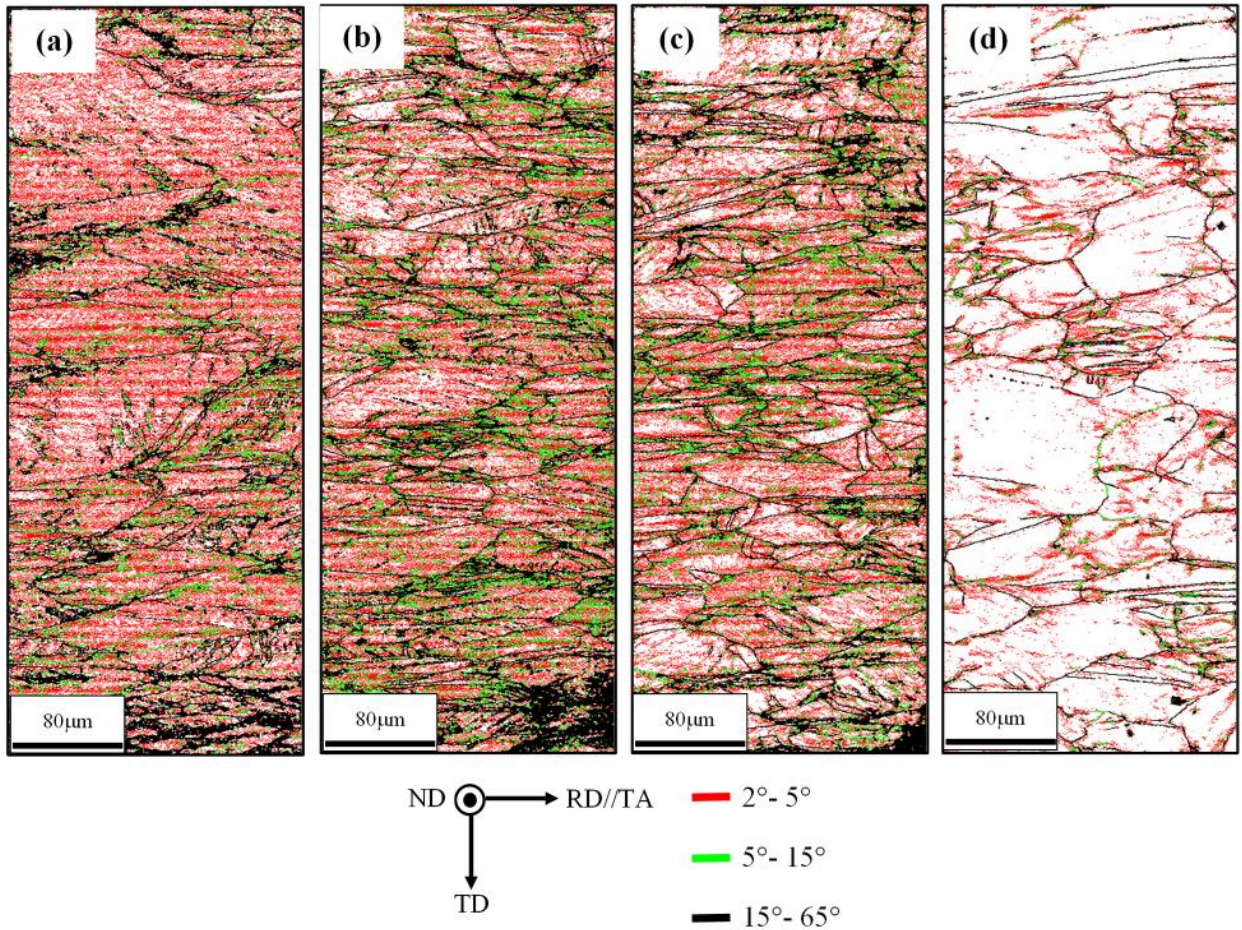

*Fig. 8.* The boundary map of fractured specimens at (a) 298 K, (b) 473 K, (c) 673 K, and (d) 873 K. The low (2-5°, 5-15°) and high angle grain boundaries (except the Σ3 twin boundaries) are delineated in red, green and black color, respectively.

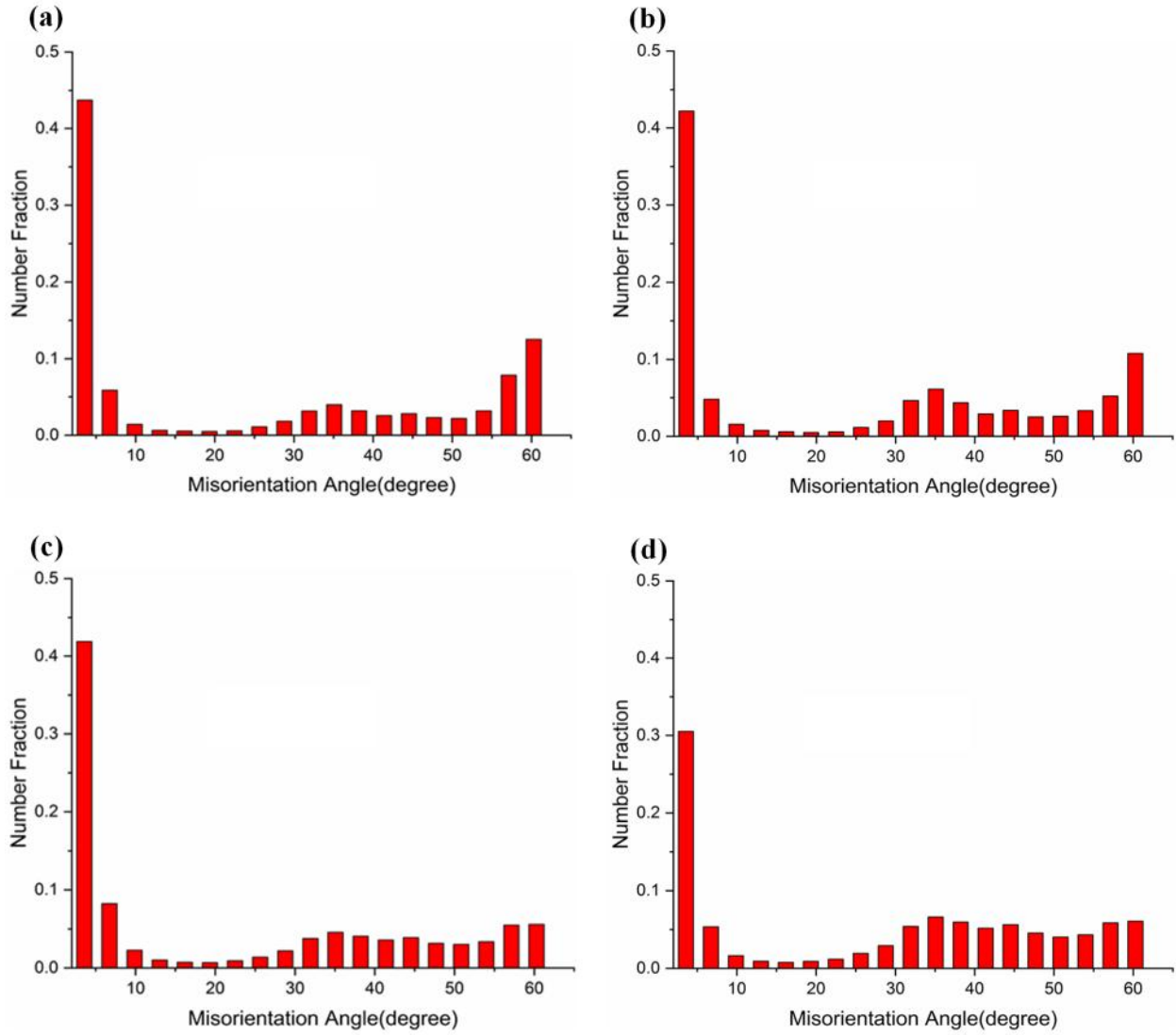

*Fig. 9.* The misorientation angle distribution of the specimens deformed at (a) 298 K, (b) 473 K, (c) 673 K and (d) 873 K.

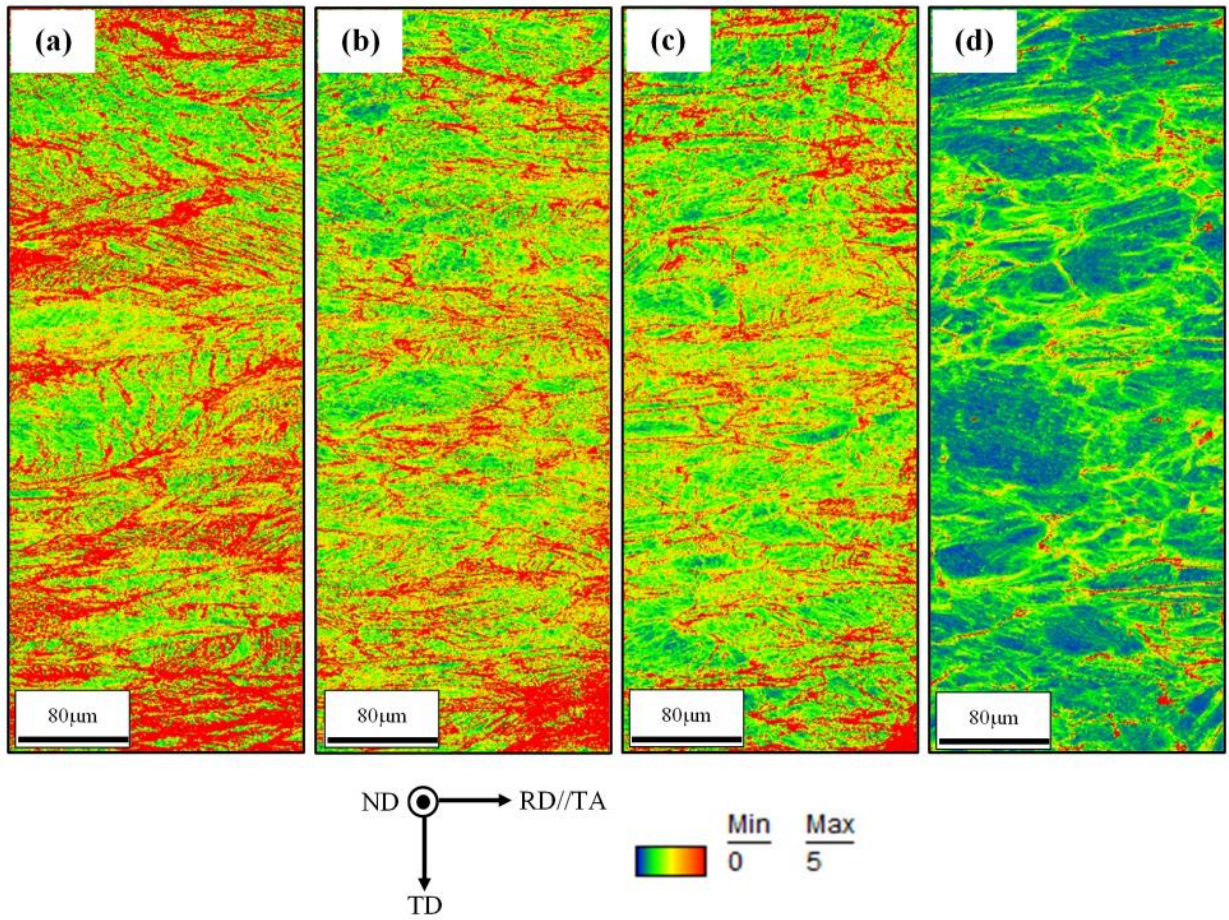

Fig. 10. The kernel average misorientation (KAM) of fractured specimens at (a) 298 K, (b) 473 K, (c) 673 K, and (d) 873 K.

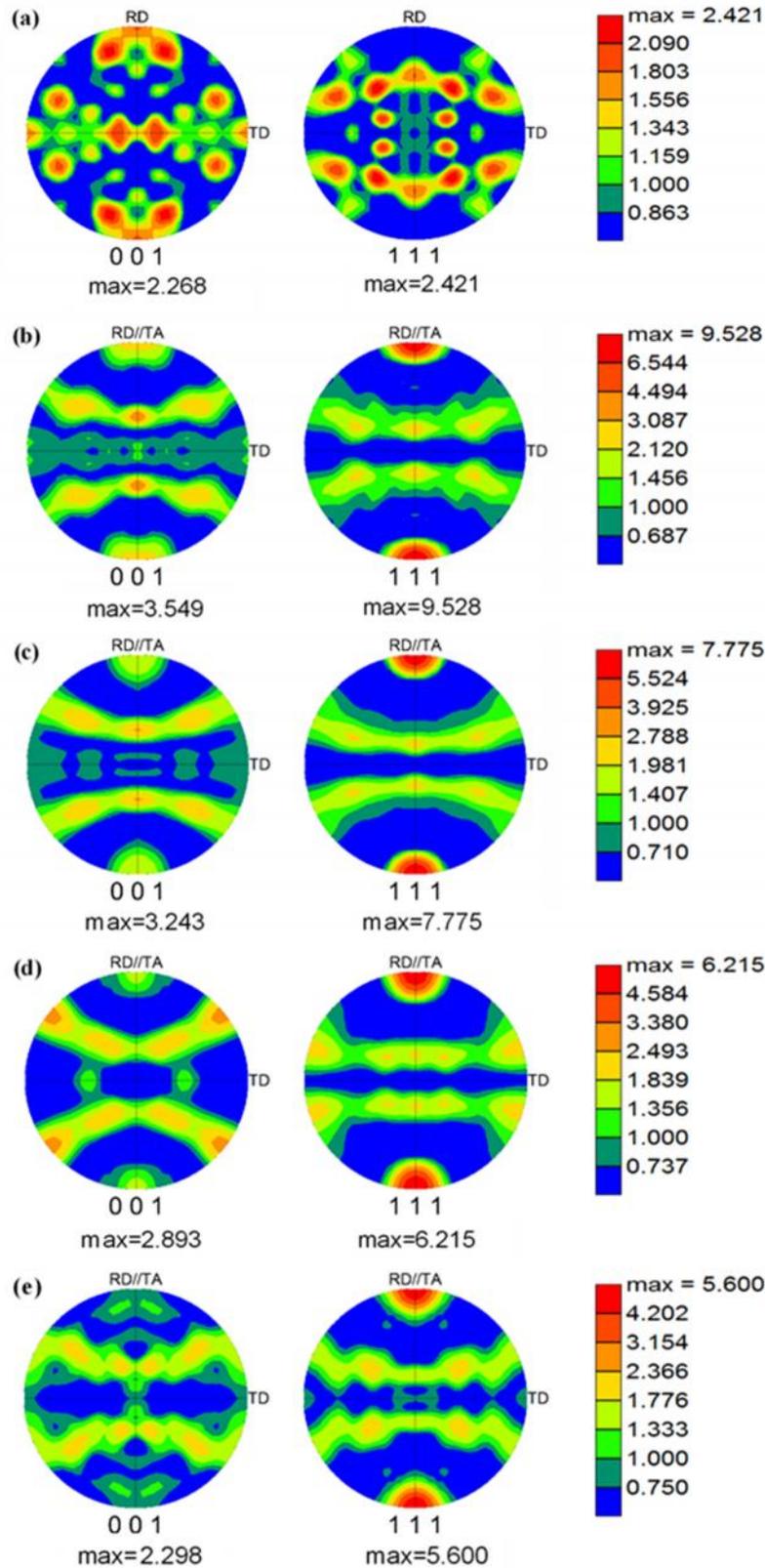

Fig. 11. Evolution of texture during deformation of investigated alloy. {001} and {111} PF of (a) initial specimen and deformed specimens at (b)298 K, (c) 473 K, (d) 673 K and (e) 873 K.

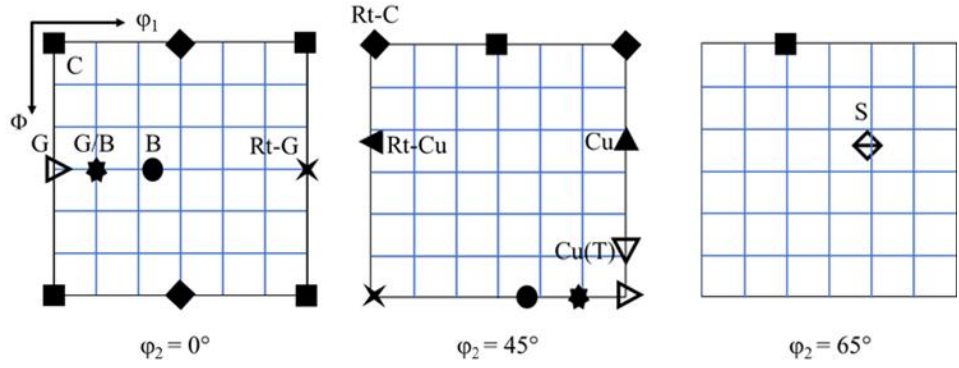

*Fig. 12. A schematic representation of the important texture components in FCC materials [60].*

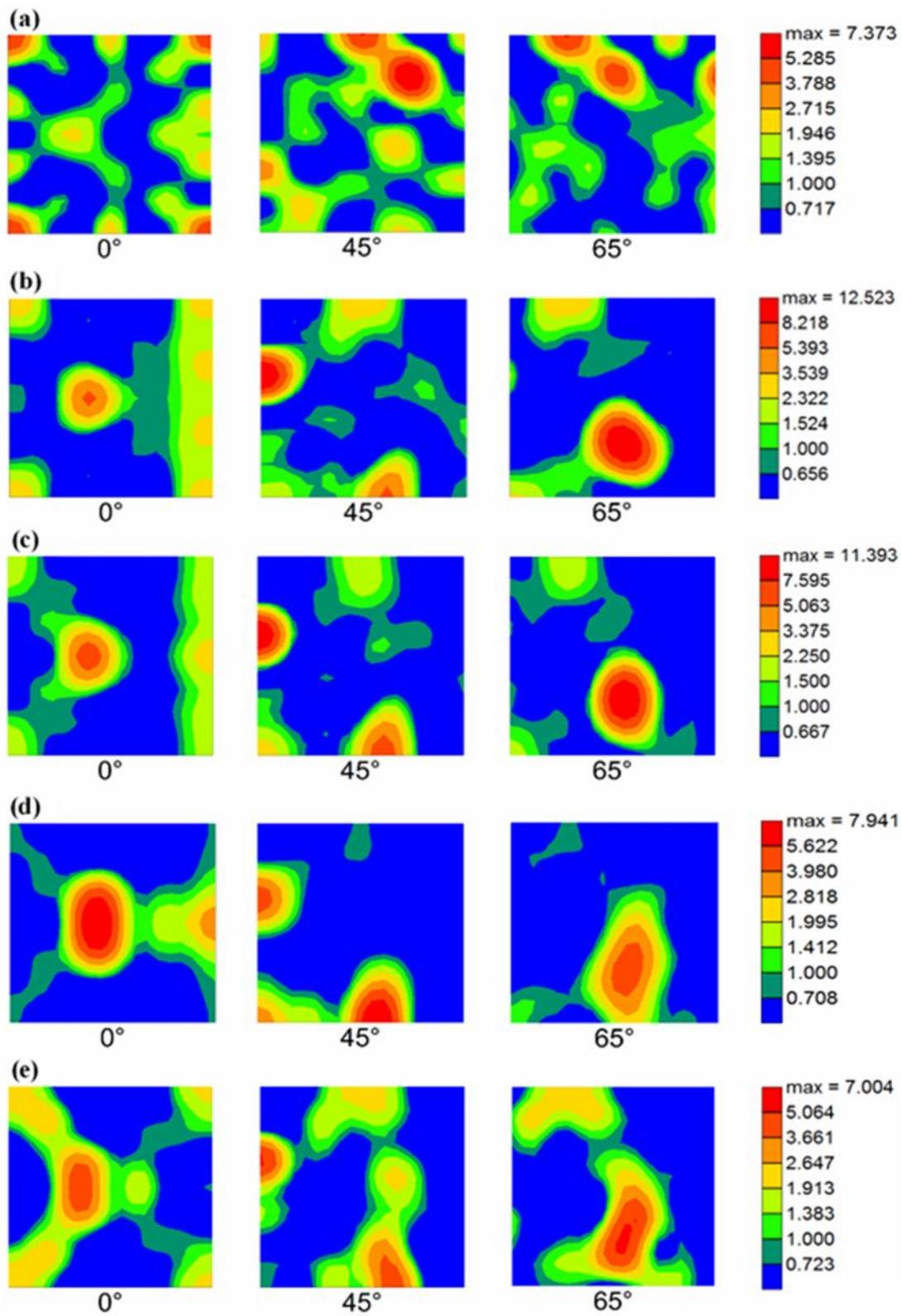

Fig. 13. The ODF sections ($\varphi_2 = 0°$, $45°$ and $65°$) obtained from the (a) initial specimen and deformed specimens at (b) 298 K, (c) 473 K, (d) 673 K and (e) 873 K.

*Table1*

*Chemical composition of the as-homogenized material measured by XRF analysis.*

| Element | Co | Cr | Fe | Mn | Ni |
|---|---|---|---|---|---|
| (at. %) | 36.73±0.2 | 19.67±0.2 | 13.08±0.2 | 16.05±0.2 | 14.47±0.2 |

*Table 2*

*Tensile properties of investigated alloy at different deformation temperatures.*

| | Deformation temperature(K) | | | | | | |
|---|---|---|---|---|---|---|---|
| | **298** | **373** | **473** | **573** | **673** | **773** | **873** |
| YS(MPa) | 227 | 202 | 187 | 175 | 162 | 157 | 146 |
| UTS(MPa) | 585 | 545 | 495 | 462 | 450 | 424 | 408 |
| UE(%) | 86 | 82 | 71 | 61 | 59 | 63 | 58 |
| TE(%) | 93 | 86 | 78 | 66 | 61 | 67 | 61 |
| n | 0.98 | 0.93 | 0.88 | 0.87 | 0.88 | 0.89 | 0.92 |
| K(MPa) | 1353 | 1257 | 1143 | 1105 | 1099 | 1044 | 1035 |

*Table 3*

*Euler angles and Miller indices for common texture components in FCC metals and alloys [60].*

| Texture component | symbol | {hkl}<uvw> | Euler angles ($\varphi_1,\Phi,\varphi_2$) | Schmid's factor | |
|---|---|---|---|---|---|
| | | | | Slip | Twinning |
| Cupper (Cu) | ▲ | {112}<111> | 90/35/45 | 0.4 | 0.4 |
| Brass (Bs) | ● | {110}<112> | 35/45/0 | 0.41 | 0.47 |
| S | ⬦ | {123}<634> | 59/37/63 | 0.46 | 0.47 |
| Goss(G) | ▷ | {110}<001> | 0/45/0 | 0.41 | 0.47 |
| Goss/Brass(G/B) | ✦ | {110}<115> | 17/45/0 | - | - |
| Cupper/Twin (Cu(T)) | ▽ | {552}<115> | 90/74/45 | - | - |
| Cube(C) | ■ | {001}<100> | 0/0/0 | 0.41 | 0.24 |
| Rotated Goss (Rt-G) | ✕ | {110}<110> | 90/45/0 | 0.41 | 0.47 |
| Rotated Cupper (Rt-Cu) | ◀ | {112}<011> | 0/35/45 | 0.27 | 0.31 |
| Rotated Cube (Rt-C) | ◆ | {001}<110> | 90/0/45 | - | - |